\documentclass[epj]{svjour}
\usepackage{graphicx}
\usepackage{dcolumn}
\usepackage{bm}

\usepackage{latexsym}
\usepackage{bm}
\usepackage{graphicx}
\usepackage[normalem]{ulem}
\usepackage{tikz}
\usetikzlibrary{positioning}
\usepackage{braket}
\usepackage{multirow}
\usepackage{dsfont}
\usepackage{mathtools}
\usepackage{xcolor}
\usepackage[normalem]{ulem}
\usepackage{cancel}
\usepackage{epsfig,dsfont}
\usepackage{subfigure}
\usepackage{bm}
\usepackage{amssymb}
\usepackage{mathrsfs}
\usepackage{amsmath}
\usepackage{enumitem}
\usepackage{dcolumn}

\newcommand{\be}{\begin{equation}}
\newcommand{\ee}{\end{equation}}
\newcommand{\bea}{\begin{eqnarray}}
\newcommand{\eea}{\end{eqnarray}}

\newcommand{\I}{\mathrm{i}}

\newcommand{\tr}{\mathrm{Tr}}

\begin{document}

\title{A finite box as a tool to distinguish free quarks
\\ from  confinement at high temperatures}
\author{L. Ya. Glozman  and C. B.~Lang
}                     
\institute{Institute of Physics,  University of Graz, A--8010 Graz, Austria}
\date{Received: date / Revised version: date}
%
\abstract{Above the pseudocritical temperature $T_c$ of chiral symmetry
restoration a chiral spin symmetry  
(a symmetry of the color charge and  of
electric confinement) emerges in QCD. This implies that QCD
is  in a confining mode and there are no free quarks.  
 At the same time
correlators of operators constrained by a conserved
current behave as if quarks were free. This explains
observed fluctuations of conserved charges and the absence
of the rho-like structures seen via dileptons.   An independent
evidence  that one is in
a confining mode is very welcome. Here we suggest  a new tool
how to distinguish free quarks from a confining mode. If
we put the system into a finite box, then if the quarks are free
one necessarily obtains a remarkable diffractive pattern in the propagator
of a conserved current. This pattern is clearly seen in a lattice calculation in
a finite box and it vanishes in the
infinite volume limit as well as in the continuum. In contrast,
the full QCD calculations in a finite box show the absence of the
diffractive pattern implying that the quarks are confined.
} 
\authorrunning{L. Ya. Glozman  and C. B.~Lang}          
\titlerunning{A finite box as a tool to distinguish free quarks
from  confinement}
\maketitle


\section{\label{sec:intro}Introduction.}

At temperatures between 100 and 200 MeV one observes in QCD
a smooth chiral symmetry restoration crossover: The quark condensate
drops from its practically zero temperature value at $T \sim 100$ MeV
to the value close to zero at  $T \sim 200$ MeV \cite{F1,K}.
Up to this crossover QCD thermodynamics is well described by a gas
of non-interacting mesons. Above the
crossover another physics regime emerges that is characterized
by a nearly perfect fluidity where there are no free quarks and gluons
and QCD is still in the confining regime.\footnote{In QCD with light
quarks only one consistent definition of confinement is known:
Confinement is the absence of color states in the spectrum. Hence deconfinement
should be accompanied by a free motion of colored quarks and gluons.} What are the physical
degrees of freedom here and how do they explain that the system is in
 the liquid regime?
 Certainly it is not yet a quark-gluon-plasma
(QGP) which is a gas and where the degrees of freedom are truly free (i.e., at most weakly interacting, not confined)
quarks and gluons.

It is clear from the lattice data for even second order quark and
baryon number fluctuations that the free quark gas limit is achieved
only at very high temperatures \cite{Bazavov}. The continuum extrapolated
screening masses for different meson channels
agree with the free quark gass estimates only at $T > 5 T_c$ for 
scalar-pseudoscalar and at $T > 3 T_c$ for vector-pseudovector channels \cite{scr}.

On the other hand there are observables  that behave as if quarks were
free particles soon above the pseudocritical
temperature of chiral symmetry restoration. 
For example, the ratio of fourth and second cumulants of quark (baryon)
number and charge fluctuations approaches a free quark gas value already
at $T \sim 200 - 250$ MeV \cite{Ejiri,Bazavov} and is considered
sometimes as "evidence of deconfinement". 
Another "evidence of deconfinement" a nonobservation of the $\rho$-like structures via dileptons
in experiments. At the same time it was established in
lattice calculations of spatial and temporal correlators \cite{R1,R2,R3} that QCD in the range $T_c - 3 T_c$ is characterized
by the chiral spin symmetry \cite{G1,G2} which is a symmetry of
the color charge and of the chromoelectric interaction\footnote{ This
symmetry was reconstructed from a large hadron spectrum degeneracy observed
on the lattice upon artificial subtraction of the near-zero modes
of the Dirac operator at zero temperature \cite{D1,D2}.}. This is
not a symmetry of the Dirac action and hence inconsistent with
free  interactionless quarks. This suggests that the degrees
of freedom are the chirally symmetric quarks bound into the
color-singlet objects by the chromoelectric field. What are
these objects?\footnote{Conditionally the regime in QCD above
$T_c$ but below $3T_c$ was named a {\em stringy fluid} to emphasize
the fact that the degrees of freedom are the ultrarelativistic chirally
symmetric quarks bound by the chromoelectric field and the
chromomagnetic effects are at least strongly suppressed.}
This symmetry was observed in lattice calculations
at zero baryon (quark) chemical potential. It should also persist
at a nonvanishing chemical potential since the quark chemical
potential in the QCD action is manifestly chiral spin symmetric \cite{G3}.

Both temporal and spatial meson correlators\footnote{Pioneering studies
 of spatial (screening) propagator are due to Ref. \cite{DeTar:1987ar}.
 In \cite{Brandt:2014uda} lattice screening masses were compared with 
 effective theory approaches.} exhibit
this symmetry very clearly at $T_c - 3T_c$, which suggests
a chromoelectric confining interaction. The screening masses extracted
from the fit with the proper $\exp( - \mu z)/z$ asymptotics  also demonstrate
clear multiplets of the chiral spin symmetry below $3 T_c$ \cite{Rohrhofer_scr}.
 At the same
time the correlators of operators constrained by a conserved
current are very close to those calculated with free quarks \cite{G5}. The latter
circumstance explains some features of fluctuations of conserved charges
above $T_c$ and why no
$\rho$-like structures are seen via dileptons. This intriguing
behavior of the correlators was a motivation for a conjecture of
a deconfinement in a $SU(2)_{color}$ subgroup\footnote{A general possibility of a deconfinement in a $SU(M)$
subgroup of $SU(N_c)$ in different large $N_c$ models was discussed
in Ref. \cite{H}.} of $SU(3)_{color}$
induced by a $SU(2)_{color}$ - $SU(2)_{isospin}$ locking \cite{G5}. 
This would explain both the chiral spin symmetry of the correlators
and at the same time their behavior in channels with conserved
currents. Because of the $SU(2)_{color}$ - $SU(2)_{isospin}$ locking
the conserved currents do not see the $SU(3)_{color}/SU(2)_{color}$
part of dynamics which is still confining. So while the correlators
of the conserved currents behave as if  quarks were free,
in reality these quarks are still in the confining mode because
of the confinement in $SU(3)_{color}/SU(2)_{color}$.

Given this intriguing situation an independent evidence
is welcome that quarks in channels with conserved currents
are still in the confining mode, even though the respective
correlators are quite close to those derived with free quarks. This question is the
subject of the present paper. We demonstrate that even if the correlators of conserved currents although
confined look like those derived for free quarks
in the continuum, we can distinguish really free quarks from these free-like
behavior by putting the system into a finite box. If quarks
are really free, in a finite box this  leads to a very specific
and bright interference pattern that does not exist in infinite volumes
or in the continuum. While we do observe such patterns in a finite box
in a free quark system, these patterns are absent in full QCD calculations
in a finite box. This  allows the conclusion that the quarks are
in a confining mode. Hence we have two independent and complementary
evidences that QCD is in the confining regime: the chiral spin
symmetry of the correlators and the absence of very pronounced interference
patterns required by free quarks in a finite box on the lattice. 

\section{\label{sec:intro}Free quarks in a finite box.}

In Minkowski space the Feynman propagator of a Dirac particle
depending on
the chronological order is either a forward running
particle ($\sim \exp(-i E t)$) or a backward running antiparticle
($\sim \exp(+i E t)$). Upon a Wick rotation to Euclidean space  the forward running
particle has an  $\sim \exp(- E t)$ dependence while the backward
running antiparticle evolves with time as  $\sim \exp(+ E t)$. If we
put the system into a finite box, e.g., on the lattice, then the rest
frame ($\mathbf{p} = 0$) time-direction propagator of a free  quark with the mass $m$ 
\begin{equation}
C_0(t) = \sum\limits_{x, y, z}
\braket{\psi(x,y,z,t)
{\bar \psi}(\mathbf{0},0)}
\label{eq:singleq}
\end{equation}
has a $C_0(t) \sim \cosh(m(t-N_t/2))$ dependence for  periodic boundary
conditions (p.b.c.) along the time direction and a
$C_0(t) \sim \sinh(m(t-N_t/2))$  form for antiperiodic boundary conditions
(a.b.c.).\footnote{On a discrete lattice $x,y,z,t$ should be discrete 
($n_x,n_y,n_z,n_t$);
$N_t$ is the lattice size in $t$-direction.} 

\begin{figure}
  \centering
  \includegraphics[scale=0.3]{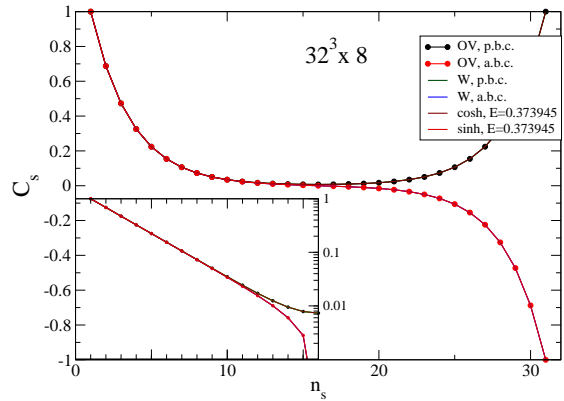} 
   \caption{ A single quark massless propagator obtained on the
   $32^3\times 8$ lattice with Wilson and overlap actions
   in comparison with the $\cosh(E (z-N_s/2))$ and $\sinh(E (z-N_s/2))$ at $E=0.37395$. Note that propagators obtained with Wilson and overlap
   actions coincide within 3 digits and cannot be distinguished in
   the plot.
}
\label{qpr}
\end{figure}

At  nonzero temperature the temporal direction becomes
short compared to the spatial one.  There are cases in
which a study of the propagators along the long spatial
direction can supply us with the information that cannot be
obtained from the temporal propagators along the short time
direction. We choose this direction to be $z$ and study the following
spatial correlators:
\begin{equation}
C_s(z) = \sum_{x, y, t}
\braket{\psi(x,y,z,t)
{\bar \psi}(\mathbf{0},0)}.
\label{eq:singleqz}
\end{equation}
This spatial single quark propagator can be straightforwardly
calculated on a finite  $N_s^3\times N_t$ lattice with
given boundary conditions. We choose antiperiodic
boundary conditions (a.b.c.) along the time direction, periodic
ones (p.b.c.) along the $x, y$ axes and either periodic or antiperiodic
along the propagation direction $z$. The results for $\tr C_s(z)$ obtained at
zero quark mass with the Wilson and overlap Dirac operators \cite{GattringerLang} are shown
in Fig. ~\ref{qpr}.
                                                              
An effective "chirally symmetric mass" $E$ for propagation of a massless
quark in $z$ direction  is very close
to the lowest Matsubara frequency $\pi/N_t = \pi/8$ and is determined
by the closest pole position $1/D(p_x=0,p_y=0,p_z=\I E,p_t=\pi/N_t)$
of the Wilson-Dirac operator in momentum space
\begin{equation}
E=\mathrm{arcosh}\left(\frac{-3+2\cos \pi/N_t}{-2+\cos \pi/N_t}\right)\;.
\label{pole}
\end{equation}
The propagator obtained
for a single quark with Wilson 
or overlap Dirac action is very accurately described by $\cosh(E (z-N_s/2))$ for p.b.c.
and by $\sinh(E (z-N_s/2))$ for a.b.c.. This  propagator can be interpreted
as a superposition of a forward ($f$) running quark with 
the "mass" $E$ and of a backward ($b$) running antiquark with the same "mass".
Symbolically the propagator can be written as
\begin{equation}
C_s(z)^{p.b.c.} \sim \exp(-Ez) + \exp(-E(N_s-z)) \equiv f + \bar b.
\label{fplusb}
\end{equation}
For the a.b.c. the propagator is
\begin{equation}
C_s(z)^{a.b.c.} \sim \exp(-Ez) - \exp(-E(N_s- z)) \equiv f - \bar b.
\label{fplusb}
\end{equation}

Having discussed the structure of a single quark propagator
in a finite box we next study propagators of quark bilinears
still keeping quarks to be noninteracting particles (I.e., due to a pure Dirac Lagrangian without any gauge fields.)
The spatial correlators of the isovector bilinear operators
$\mathcal{O}_\Gamma(x,y,z,t) =\bar \psi(x,y,z,t) \Gamma\, \frac{\vec \tau}{2}\, \psi(x,y,z,t)$ with $\Gamma$ being out of a set of $\gamma$-matrices are
\begin{equation}
C_\Gamma(z) = \sum_{x, y, t}
\braket{\mathcal{O}_\Gamma(x,y,z,t)
\mathcal{O}_\Gamma(\mathbf{0},0)^\dagger}.
\label{eq:momentumprojection}
\end{equation} 
The isovector fermion bilinears are named
according to Table \ref{tab:ops}. 

\begin{table*}
\center
\begin{tabular}{cccll}
\hline\hline
\rule{0pt}{3ex}
 Name        &
 Dirac structure  $\Gamma$ &
 \quad Abbreviation  \quad &
 \multicolumn{2}{l}{
 } 
 
\\[1ex]
\hline
\rule{0pt}{3ex}

\textit{Pseudoscalar}        & $\gamma_5$                 & $PS$         &
 \multirow{2}{1cm}{$\left.\begin{aligned}\\ \end{aligned}\right] ~~U(1)_A$}& \\
\textit{Scalar}              & $\mathds{1}$               & $S$        &  & 
\\[1ex]
\hline
\rule{0pt}{3ex}
\textit{Axial-vector}        & $\gamma_k\gamma_5$         & $\mathbf{A}$ & 
\multirow{2}{1cm}{$\left.\begin{aligned}\\ \end{aligned}\right] ~~SU(2)_A$}&\\
\textit{Vector}              & $\gamma_k$                 & $\mathbf{V}$ & & \\
\textit{Tensor-vector}       & $\gamma_k\gamma_3$         & $\mathbf{T}$ & 
\multirow{2}{1cm}{$\left.\begin{aligned}\\ \end{aligned}\right] ~~U(1)_A$} &\\
\textit{Axial-tensor-vector} & $\gamma_k\gamma_3\gamma_5$ & $\mathbf{X}$ & &\\[1ex]
\hline\hline
\end{tabular}
\caption{Fermion isovector bilinears and their $U(1)_A$ and $SU(2)_L \times SU(2)_R$ transformation properties
(last column). This classification assumes propagation in $z$-direction. The
open vector index $k$ here runs over the components $1,2,4$, i.e., $x,y$ and $t$.}
\label{tab:ops}
\end{table*}

A complete set of such propagators  in the continuum (in infinite volume)
has been determined analytically in Ref. \cite{R2}. There
these correlators are given as superpositions of the decaying exponents
 $\exp(-2\pi n_z/N_t)/(2\pi n_z/ N_t )$,
$\exp(-2\pi n_z/ N_t)/(2\pi n_z/N_t )^2, \ldots$ and terms
with higher Matsubara frequencies and represent the propagators of the
forward propagating "mesons" that are made from noninteracting quarks.  

In a finite box a quark propagator of a given flavor 
is represented as a sum (for p.b.c) or difference (for a.b.c.)
of the forward propagating quark and of the backward propagating
antiquark. The same is true for the antiquark propagator,
that is a sum (or difference) of the forward propagating antiquark
and of the backward propagating quark. Consequently
correlators of the  bilinears should be superpositions
of four terms:
\bea
p.b.c.:&~&  (f + \bar b)(\bar f + b)=f \bar f+b\bar b+f b+\bar b \bar f\;,
\label{per}\\
a.b.c.:&~& (f - \bar b)(\bar f - b)=f \bar f+b\bar b-f b-\bar b \bar f\;.
\label{aper}
\eea	      
Note that the two terms $\sim \bar f f$ and $\sim \bar b b$
represent the forward and backward propagating meson-like
system. The other two terms $\sim  f  b$ and $\sim \bar f \bar b$  
do not represent any meson-like system. More precisely, they represent
a quark-antiquark system where the quark and the antiquark
are shifted relative to each other by a large distance $N_s$. These terms are necessarily
present in the correlators of the quark-antiquark bilinears if
quarks are free particles that do not interact. They exist
only in a finite box and vanish in the infinite volume limit or
in the physical continuum. If we put the system of free quarks
into a finite box, then these "unphysical" terms must be observable
since they interfere with the "physical" meson-like amplitudes.
The interference should be clearly seen in cases when the
 "physical" and  "unphysical" terms are of a similar magnitude
 and interfere destructively. Since the "unphysical" terms are very
 small one should expect this destructive interference to be
 clearly visible only when the "physical" terms are also very small.
 The numerical results for the  propagators calculated with
 free noninteracting quarks \cite{R1} show that the largest slope 
 of the decay takes place with the operators $V_t, A_t, T_x, T_y, X_x, X_y$ 
 and all  other operators $V_x, V_y, A_x, A_y, ...$ have 
 smaller decay rate \footnote{$V_t$ refers to the $\bar \psi(x,y,z,t) \gamma_4 \, \frac{\vec \tau}{2}\, \psi(x,y,z,t)$ operator, etc, see the legend to the
 Table.}. This suggests that the "physical"
 meson-like amplitude becomes sufficiently small at large $z$ for
 the operators $V_t ,A_t, T_x, T_y, X_x, X_y$ and we can expect in this case
 well visible interference effects of the "physical" and "unphysical"
 amplitudes.   

\begin{figure}
  \centering
  \includegraphics[scale=0.3]{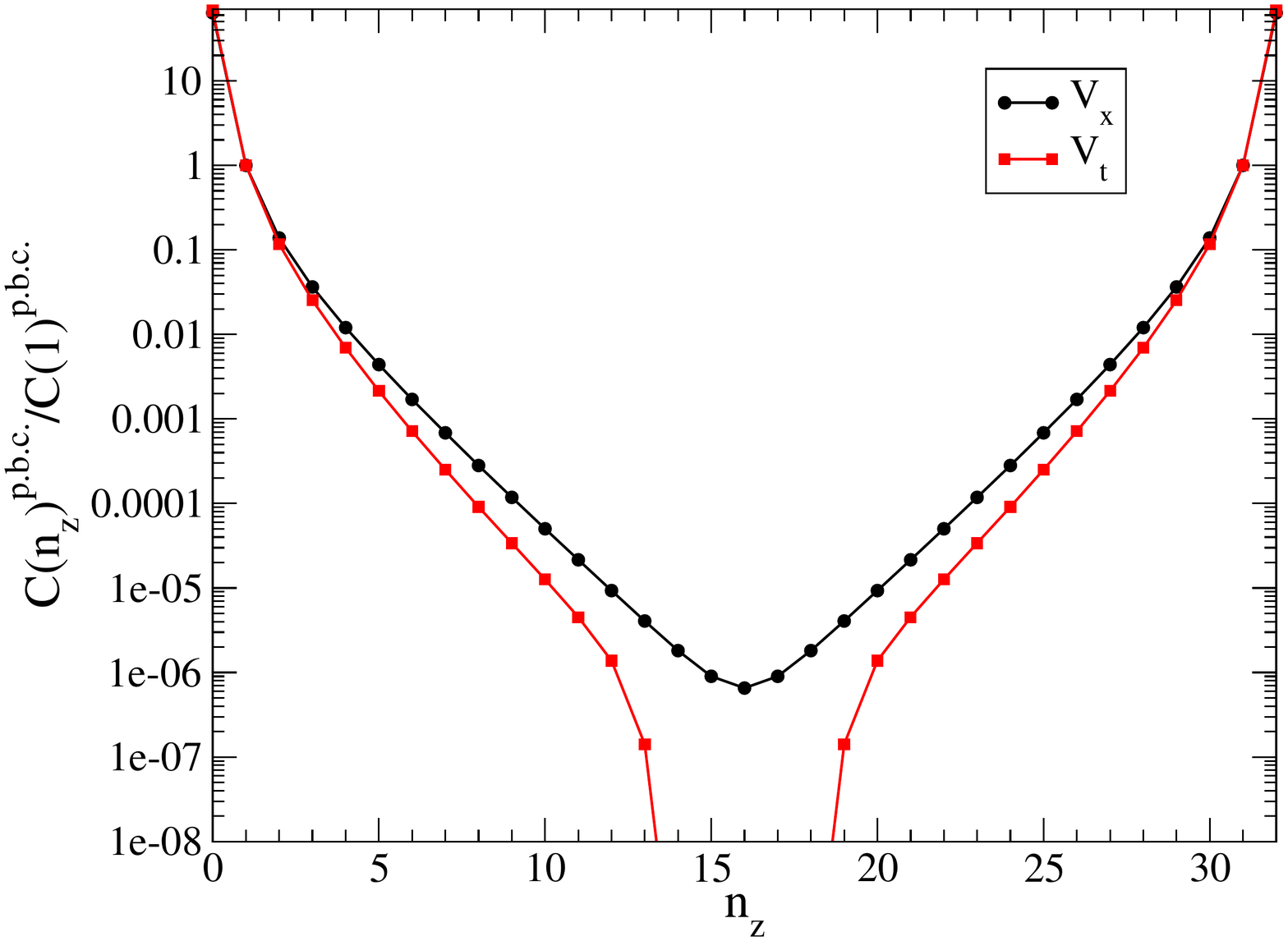} 
   \includegraphics[scale=0.3]{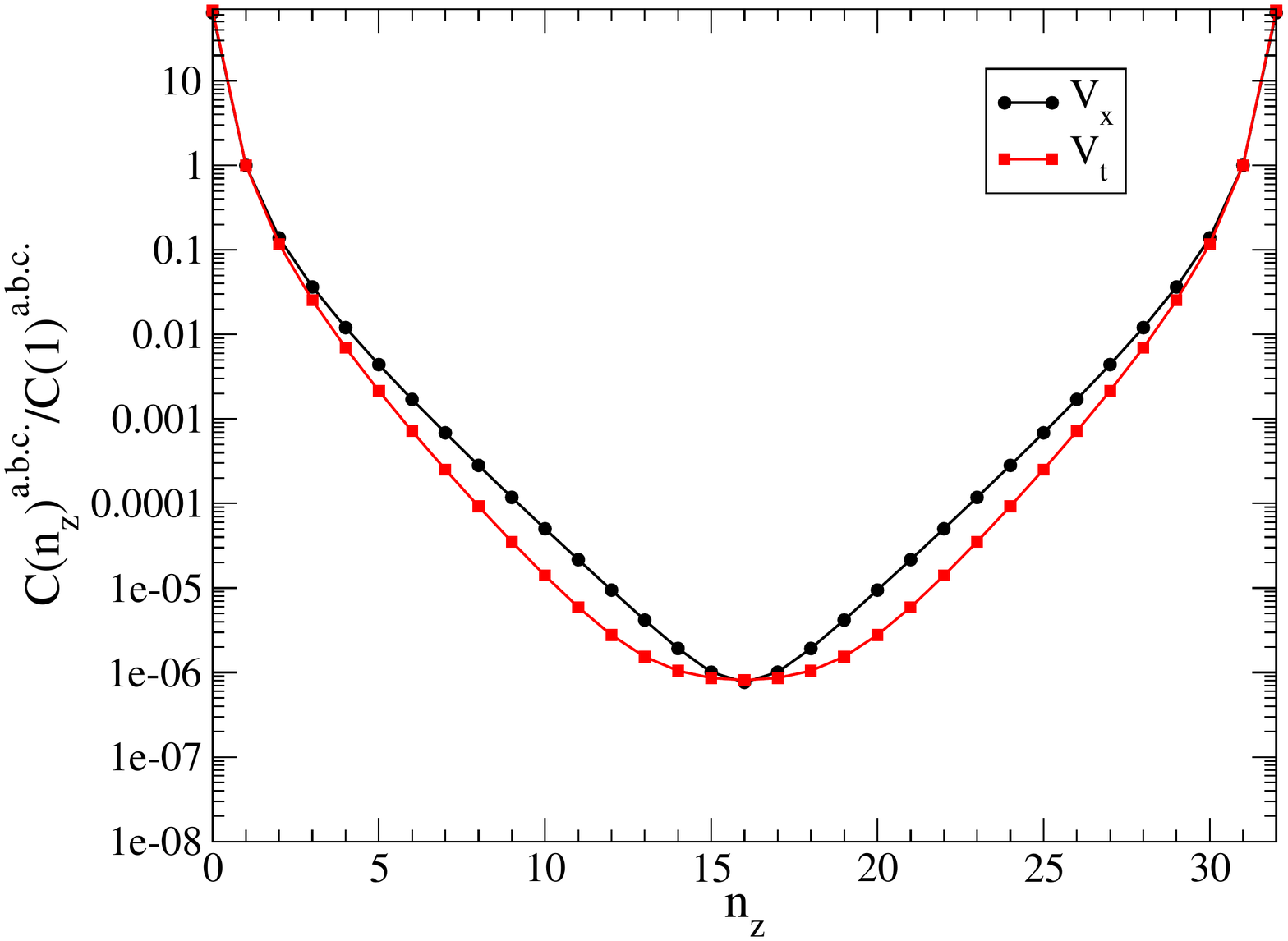}
   \caption{ Correlators of the $V_t$ and $V_x$ bilinears
   on a $32^3\times 8$ lattice with the  overlap action
   with periodic (p.b.c., left panel) and antiperiodic (a.b.c., right
   panel) boundary conditions in $z$ direction.
   The $V_t$ correlator in the left panel is negative for $n_z \sim 14 - 18$.
   The correlators are normalized to 1 at $n_z=1$.
}
\label{VxVt}
\end{figure}
The correlators calculated with the overlap action on the 
$ 32^3 \times 8$ lattice with the $V_t$ operator are shown
in Fig. \ref{VxVt}. The correlators of the $A_t,T_x,T_y,X_x,X_y$
operators are similar. We also show in the same figure  the
correlators of the $V_x$ operator that demonstrates a
smaller decay rate.

We clearly see a typical diffractive structure for
the correlator of the $V_t$ operator at large $z$
and when  p.b.c. are imposed the correlator becomes negative for $z \sim 14 - 18$. 
This was first noted in Refs. \cite{I,R1} but remained unexplained.
Now we realize that this structure is the result of the destructive
interference of the "physical" and "unphysical" amplitudes. It is
an immanent property of a system of free quarks in a finite box.
In contrast, the correlator of the $V_x$ operator does not
show a diffractive structure because the "physical" terms
in this case are always essentially larger than the "unphysical" ones.

How to check this picture of the destructive interference?
If we change from p.b.c. to  a.b.c. one should expect
a constructive interference of the "physical" and "unphysical"
terms. Hence the diffractive structure should disappear. This is
precisely what happens. 

Numerical checks indicate that the diffractive structure
disappears exponentially upon increase of $N_s$ (at fixed $N_t$).
Hence it vanishes both in large lattice volumes as well
as in the continuum theory.

\section{\label{sec:3}Comparison of the full QCD and free quarks correlators
in a finite box.}

We have established in the previous section that if quarks are
free, then the spatial correlators of the conserved currents $V_t,A_t$ and 
of some other operators exhibit on a finite lattice remarkable diffractive patterns. These
are a consequence of the fact that for free quarks there are necessarily
amplitudes that represent a "meson-like" propagation, called
"physical", and "unphysical" amplitudes that do not correspond
to any meson-like system. These "physical" and "unphysical"
amplitudes interfere destructively. The "unphysical" amplitudes vanish on
the infinite lattice as well as in the continuum and the diffractive
pattern disappears.

 At the same time the "unphysical" terms are much
smaller than the "physical" ones for another set of operators and
the diffractive pattern does not exist. These features are a solid
prediction of a free quark system put on a finite lattice.

In the continuum full QCD above $T_c$ the spatial and temporal correlators of the
conserved currents are close to those ones calculated with noninteracting quarks \cite{G5}.
In reality quarks cannot be  free since these correlators are subject to the
chiral spin symmetry that is not a symmetry of the Dirac action.
Is there another means to decide that the quarks are not free?
The answer is affirmative. When we solve  QCD at high temperatures
on the finite lattice if the quarks are not confined (i.e., free),
one should observe the diffractive pattern as described above. If such a
pattern is missing, then we could safely conclude that the quarks are
not free. This is demonstrated below.

\begin{figure}
  \centering
  \includegraphics[scale=0.45]{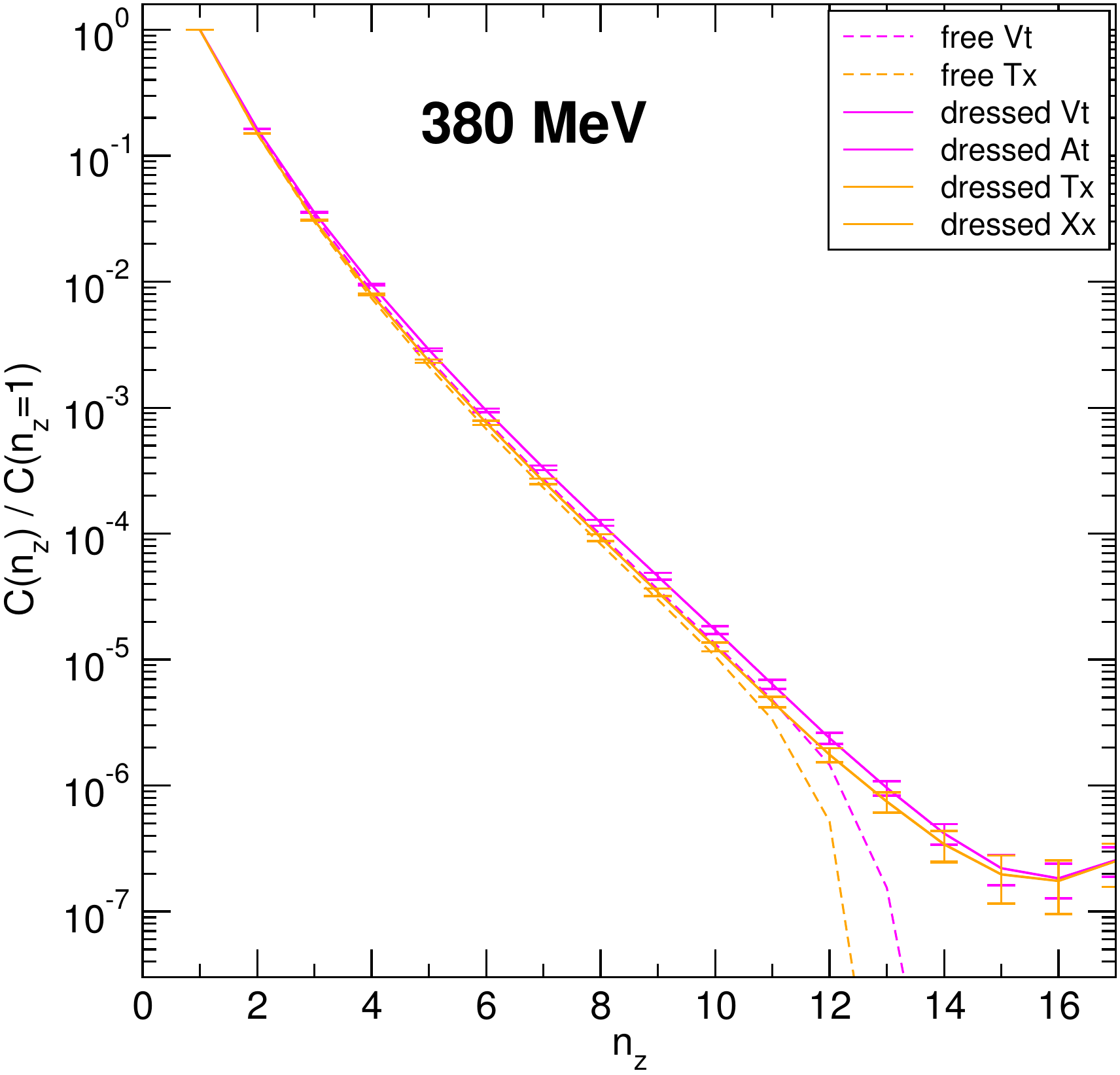}
  \caption{Correlators of the $ V_t,  A_t,  T_x, X_x$ operators
     in full QCD at $T$= 380~MeV ($\sim 2.2 T _c$) for $32^3 \times 8$
     lattice (abbreviated as \textit{dressed})
      and with non-interacting quarks (\textit{free}) on the same lattice. From Ref. \cite{R1}. }
  \label{fig:e3}
\end{figure}

In Fig. \ref{fig:e3} we show correlators normalized to 1 at $n_z=1$ 
 built with
the  $V_t, A_t, T_x, X_x$ operators calculated in $N_F=2$ QCD  with the domain wall
Dirac operator at
physical quark masses on $32^3 \times 8$ lattice at $T=380$ MeV ($2.2 T_c$)
\cite{R1}. The boundary conditions for quarks are a.b.c. in
time direction and p.b.c. in all spatial directions.
The solid curves represent the full QCD results while the dashed curves
are correlators calculated on the same lattice with the same Dirac operator
with free noninteracting quarks, i.e.  computed with a trivial gauge field configuration ($U=1$).
The free quark correlator of the $V_t$ operator corresponds to the
results shown in Fig. \ref{VxVt}. It is rather obvious that
the free quark results obtained with the domain wall Dirac operator 
in Fig. \ref{fig:e3} are similar to those obtained with the overlap Dirac
operator in Fig. \ref{VxVt}. In both cases we see a remarkable diffractive
structure around $n_z \sim 12 - 20$. 

\begin{figure}
  \centering
  \includegraphics[scale=0.3]{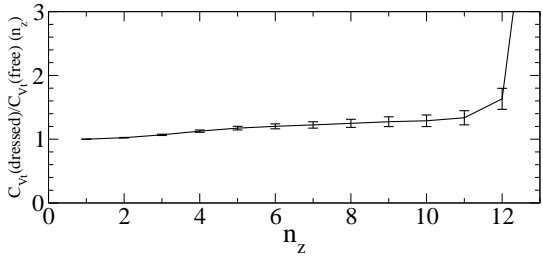}
  \caption{Ratio of  correlator of   $ V_t$ operator
     in full QCD at $T$= 380~MeV ($\sim 2.2 T_c$) for $32^3 \times 8$
     lattice (abbreviated as \textit{dressed}) to $ V_t$ correlator 
      calculated  with non-interacting quarks on the same lattice size (\textit{free}).
  }
  \label{fig:4}
\end{figure}

This diffractive structure is
induced via a destructive interference of the "physical" meson-like 
amplitudes with the "unphysical" amplitudes that do not correspond
to any meson-like system. This structure necessarily exists on a finite
lattice if the quarks are free noninteracting particles. In contrast,
the full QCD results do not show this diffractive pattern.
The diffractive pattern is induced by the deep
infrared region, i.e., by large distances between quarks.
Even though the propagator of the conserved
current $V_t$ in full QCD is rather close to the free quarks 
propagator at $ n_z < 11 $, for the ratio of both propagators see Fig. \ref{fig:4},
it does not represent a free quark system
but describes a propagation of a meson-like system with confinement.
All "unphysical" terms that exist in the case of the free quark system
are killed by a confining gluonic interaction between quarks that are
separated by a large spatial distance. 
The fluctuations of conserved charges are given by  integrals
of the correlator. Since the crucial deviations of the free quark
correlator from the QCD correlator are seen only at very large
distances (where the absolute value of the correlators is suppressed by
5-6 orders of magnitude) a sensitivity of the fluctuations of the
conserved charge to the deep infrared is only weak.

\begin{figure}
  \centering
  \includegraphics[scale=0.45]{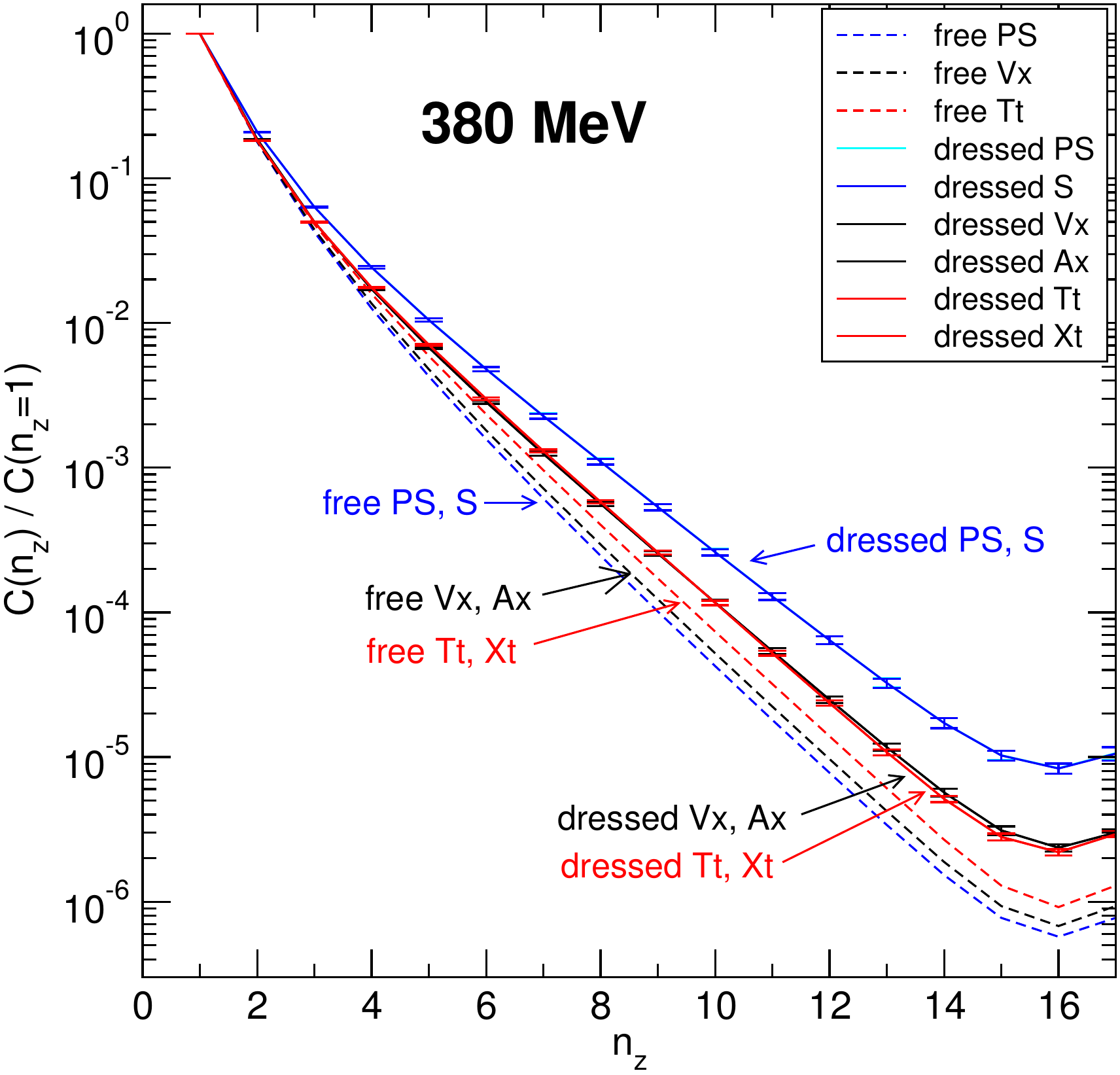}
  \caption{Correlators of the $PS, S, V_x, A_x, T_t, X_t$ operators
     in full QCD at $T$= 380~MeV ($\sim 2.2 T_c$) for $32^3 \times 8$
     lattice (abbreviated as \textit{dressed})
      and with non-interacting quarks (\textit{free}) on the same lattice. From Ref. \cite{R1}.
  }
  \label{fig:e2}
\end{figure}

The propagator of the $V_x$ operator, that is not constrained by a current
conservation, demonstrates the absence of the diffractive structure
both in full QCD as well as for free quarks, see Fig. \ref{fig:e2}.

We summarize this section with the principal result of the
present paper. There are two independent evidences that
a system with quantum numbers of a conserved current is in
a confining mode above $T_c$. The first evidence are  the very clear
patterns of the chiral spin symmetry both in spatial and temporal
correlators \cite{R1,R2,R3}. The second evidence, demonstrated
in the present paper, is the absence of the diffractive pattern
required by a system of free quarks.

\section{\label{sec:4}Discussion and conclusions.}

We have demonstrated that on a finite lattice in a system
of free noninteracting quarks the spatial propagators of the
bilinear quark-antiquark operators exhibit 
in  case of periodic boundary conditions
along the propagation direction 
a  diffractive pattern  for  operators that are constrained by
a current conservation
and for some other operators. This
diffractive pattern is a consequence of a destructive interference
of the amplitudes that correspond to the propagation of a meson-like
system made of a quark and an antiquark with amplitudes that do not describe
any meson-like system. The latter amplitudes arise exclusively due
to a finiteness of a box and vanish on an infinite lattice or in the
continuum. The latter amplitudes as well as a diffractive pattern is an
immanent property of the free quark system in a finite lattice.

In QCD the correlators of conserved currents  above a
chiral symmetry restoration crossover behave as if quarks were
free, i.e. the correlators of these currents calculated in QCD
are rather close to correlators obtained with free noninteracting quarks \cite{G5}.
This explains why some cumulants of fluctuations of conserved charges suggest
 a free
quark gas-like behavior very soon above $T_c$ as well as
absence of the rho-like structures observed via dileptons in heavy
ion collisions. At the same time these correlators as well as another
ones are a subject to a chiral spin symmetry \cite{G1,G2} at $T_c - 3 T_c$ \cite{R1,R2,R3}. 
This symmetry is not a symmetry of the Dirac action and hence inconsistent with free
noninteracting quarks. It is a symmetry
of the color charge in QCD and  it indicates that QCD is in the confining regime
where the chromoelectric interaction binds the chirally symmetric quarks
into color-singlet objects ("strings") and a contribution of the
chromomagnetic interaction is at least strongly suppressed. 

An independent evidence  confirming
that the quark-antiquark systems with a conserved current quantum numbers
are indeed in the confining regime is supplied by QCD  
on the lattice in a finite box. If the quarks are  free, then there must 
be a diffractive pattern described above that is induced by quarks that
are separated by a large space distance. In full QCD calculations above $T_c$ in a
finite box such pattern is not observed. It follows then that the
quarks are not free and confining chromoelectric dynamics kills all
amplitudes that do not correspond to propagating mesons.

 Hence we have two independent and complementary evidences of
 confinement in  at $T_c - 3 T_c$. These are the
 chiral spin symmetry of correlators and the absence of a diffractive
 structure required by free quarks in a finite 
 box. This regime we
 have conditionally called "stringy fluid"  \cite{G3,R2}.
  
 At temperatures above $3 T_c$ the chiral spin symmetry smoothly disappears  \cite{R2}
 and correlators of all operators  approach correlators calculated
 with free quarks. This suggests that eventually the color charge 
 and electric confining interaction is Debye
 screened within  $SU(3)_{color}$ (cf., Ref.\cite{Kajantie,Ko}). Still the correlators of the conserved
 currents in a finite box do not show the diffractive structure required
 by really free quarks \cite{R2}. This indicates that there are no  free, noninteracting
 quarks and the system is still in the confining regime (defining confinement as the absence of free quarks and gluons.) The latter fact
 can be explained by the presence of a weak magnetic confinement
 at very high temperatures. It is known that at very high temperatures
 QCD is dimensionally reduced to a weakly coupled 3-dimensional pure magnetic
 theory \cite{AP}. Even though the theory is weakly coupled, there
 is a pure magnetic weak "confining" interaction that does not
 allow quarks to be completely free \cite{mag,Ko,F}. In this regime
 all properties of QCD should be close to the quark-gluon-plasma
 regime.
 
 We emphasize that the absence of a diffractive pattern
 in full QCD spatial correlators evidences confinement, but it
 cannot distinguish between the electric and magnetic confinement.
 Only the chiral spin symmetry observed in the range $T_c -  3 T_c$
 \cite{R2,R3} does suggest that it is an electric confinement that drives properties
 of QCD in the stringy fluid regime.

\bigskip
We  thank T. Cohen, C. Gattringer, O. Philipsen  and R. Pisarski 
for careful  reading of the ms.


\end{document}